\documentclass[aps,pre,twocolumn,groupedaddress]{revtex4}
\usepackage{graphicx}
\usepackage{amsmath}
\usepackage{amssymb}

\begin{document}


\title{Multicritical points for the spin glass models on hierarchical lattices}


\author{Masayuki Ohzeki$^{1}$, Hidetoshi Nishimori$^{1}$, and A. Nihat Berker$^{2,3,4}$}

\affiliation{$^{1}$Department of Physics, Tokyo Institute of Technology, Oh-okayama, Meguro-ku, Tokyo 152-8551, Japan\\
$^{2}$College of Sciences and Arts, Ko\c{c} University, Sar\i yer 34450, Istanbul, Turkey\\
$^{3}$Department of Physics, Massachusetts Institute of Technology, Cambridge, Massachusetts 02139, U.S.A. and\\
$^4$Feza G\"ursey Research Institute, T\"UB\.ITAK - Bosphorus University, \c{C}engelk\"oy 34684, Istanbul, Turkey}


\date{\today}

\begin{abstract}
The locations of multicritical points on many hierarchical lattices are numerically investigated by the renormalization group analysis. 
The results are compared with an analytical conjecture derived by using the duality, the gauge symmetry and the replica method.
We find that the conjecture does not give the exact answer but leads to locations slightly away from the numerically reliable data.
We propose an improved conjecture to give more precise predictions of the multicritical points than the conventional one.
This improvement is inspired by a new point of view coming from renormalization group and succeeds in deriving very consistent answers with many numerical data.
\end{abstract}


\maketitle
\section{Introduction}
The problem of spin glass, which is one of the most challenging subjects in statistical physics, has been analyzed extensively by the mean field theory but has not been sufficiently understood for finite dimensional systems.\cite{EA,SK,Young}
Most approaches to the difficult problem of finite dimensional spin glasses rely on approximate techniques, numerical simulations and phenomenological theories.\cite{Young}

A lot of important facts have been established by such approaches to finite dimensional spin glasses.
Nevertheless, it is very important to derive exact or rigorous results to check validity of approximate approaches.
A useful method along this line is the gauge theory, which enables us to find a special subspace of a phase diagram for spin glass models known as the Nishimori line.\cite{HN81,HNbook}
In this subspace, we can calculate the exact value of the internal energy and evaluate the upper bound of the specific heat.
It is shown rigorously that the Nishimori line runs through the ferromagnetic phase and the paramagnetic phase.
Moreover it is expected that the multicritical point, where the phase boundaries between the spin glass phase, paramagnetic phase, and ferromagnetic phase merge, is located on the Nishimori line.\cite{DoussalHarris}

One of the recent developments using the gauge theory is a conjecture of the exact location of the multicritical point for spin glasses, especially in two dimensions.\cite{NN,MNN}
The prediction is very close to numerical results,\cite{Nstat} and is considered to be very useful in the analysis of numerical data for critical exponents.

The essential part to derive the conjecture consists of duality and the replica method as explained below.
Duality is a useful tool to obtain the exact location of the transition point for spin systems without disorder.
Because the spin glass models have quenched disorder, we cannot directly apply the duality to spin glass models.
Nevertheless, by the replica method, the problem reduces to non-random systems to which we can apply duality.
We then assume that a single relation gives the multicritical point similarly to the case of the duality relation on the transition point for the pure Ising model in two dimensions.
This is one of the hypotheses on the conjecture.
In addition we expect that the above-mentioned relation for the multicritical point is satisfied even when the replica number goes to zero to analyze systems with quenched randomness.

The validity of assumptions to derive the conjectures as mentioned above has not been established rigorously.
Nevertheless the conjecture has given predictions very close to many independent numerical results.\cite{NN,MNN,Nstat,TSN,NO}
However, Hinczewski and Berker found several examples by the exact renormalization analysis in which, especially for the hierarchical lattices, the conjecture did not give good predictions.\cite{HB}
It is expected generally that the renormalization group analysis on hierarchical lattices gives exact results.
Therefore such discrepancies found by the renormalization group analysis on hierarchical lattice should be taken seriously for the conjecture even though the amount of discrepancies is small.
If these discrepancies are genuine, we have to consider the reason why there are cases for which the conjecture does not work well.
Conversely, we would like to know why the conjecture always gives accurate results if not exact.
It is also desirable to improve the conjecture to predict more precise location of the multicritical point.

These observations have given us motivations to investigate many more hierarchical lattices to check if the locations of the multicritical points are away from predictions by the conjecture.
We employ the technique by Nobre\cite{Nobre} to examine the phase transitions and to estimate the locations of the multicritical points on several hierarchical lattices.
Then, we discuss reasons why there are such discrepancies for the cases of hierarchical lattices and propose a method to improve the conjecture to derive more precise locations of multicritical points.

The presented paper is organized as follows.
In Sec. II, we review previous results for the location of the multicritical point, and show some examples of the discrepancies between the conjecture and the numerical results.
We explain the properties of hierarchical lattices in Sec. III.
In Sec. IV, we explain Nobre's method to examine the phase transition and then estimate the values of the locations of the multicritical points for several cases. 
We find some discrepancies between the conjecture and the numerically estimated results here.
Therefore we have to consider the reason why there are some cases that the conjecture does not work well.
The conjecture relies on the replica method, and we assume that the analytical continuation of the replica number to zero does not cause troubles as in most cases studied so far.
Therefore we investigate the replicated systems for the $\pm J$ Ising model in Sec. V.
Then we consider improvement of the conjecture in Sec. VI and show successful results to predict more precise locations of the multicritical points in Sec. VII.
In Sec. VIII, discussions and future outlook are given.
\section{Model and Conjecture}
We study the random-bond Ising model, defined by the Hamiltonian,
\begin{equation}
H = - \sum_{\langle ij \rangle} J_{ij} \sigma_{i}\sigma_{j}
\end{equation}
where $\sigma_i$ is the Ising spin taking values $\pm 1$, and $J_{ij}$ denotes the quenched random coupling.
In this paper, we consider two types of distribution functions for $J_{ij}$, the $\pm J$ model and the Gaussian model.

For this random-bond Ising model on two-dimensional lattices, a method to predict the precise location of multicritical point has been proposed.\cite{NN,MNN,Nstat,TSN,NO}
This method relies on the duality and the replica method applied to spin glass models with gauge symmetry.
It has been considered to be a conjecture for the exact location of the multicritical point for systems satisfying certain conditions like self duality.
According to this conjecture, the exact location of the multicritical point for the $\pm J$ Ising model on the square lattice is determined by a single equation as follows,
\begin{equation}
 -p \log_2 p -(1-p)\log_2(1-p) = \frac{1}{2},\label{conjecture}
\end{equation}
where $p$ is the probability of $J_{ij} = J>0$ for the $\pm J$ Ising model.
The left-hand side of this equation is the binary entropy and will be written as $H(p)$.
We obtain the value $p_c=0.889972$ ($\approx 0.8900$), solving this equation.
This result is in reasonable agreement with existing numerical results as shown in Table \ref{Comparison}.
\begin{table}
\begin{tabular}{ccc}
\hline
Type & Conjecture & Numerical result \\
\hline
SQ $\pm J$ & $p_c = 0.889972$\cite{NN,MNN} & $0.8900(5)$\cite{Qeiroz}\\
 & & $0.8894(9)$\cite{pm2} \\
 & & $0.8907(2)$\cite{pm3} \\
 & & $0.8906(2)$\cite{pm4} \\
 & & $0.8905(5)$\cite{pm5} \\
 & & $0.8907(4)$\cite{Gaussian} \\
SQ Gaussian & $J_0/J^2 = 1.021770$\cite{NN,MNN} & $1.02098(4)$\cite{Gaussian} \\
TR $\pm J$ & $p_c = 0.835806$\cite{NO} & $0.8355(5)$\cite{Qeiroz} \\
HEX $\pm J$ & $p_c = 0.932704$\cite{NO} & $0.9325(5)$\cite{Qeiroz} \\
\hline
\end{tabular}
\caption{{\small 
Comparisons between the conjectured values and the numerical results. SQ denotes the square lattice, TR means the triangular lattice, and HEX expresses the hexagonal lattice.}}\label{Comparison}
\end{table}
It is also possible to obtain the location of the multicritical point for the Gaussian model with the average $J_0$ and the variance $J^2$ from the following equation,\cite{NN,MNN}
\begin{equation}
 \int^{\infty}_{-\infty} dJ_{ij} P(J_{ij})\log_2\left\{1+\exp(-2\beta J_{ij})\right\} = \frac{1}{2},\label{Gconjecture}
\end{equation}
where $P(J_{ij})$ is the Gaussian distribution function, and $\beta$ satisfies the condition of the Nishimori line $\beta =J_0/J^2$. 
We will write the left-hand side of this equation as $H_G(J_0/J^2)$.
The solution of Eq. (\ref{Gconjecture}) for the Gaussian model is $J_0/J^2 = 1.021770$, which is also compared with the existing numerical result in Table \ref{Comparison}.
It is not easy to determine from these data whether the conjecture actually gives the exact result.

Equation (\ref{conjecture}) applies also to models defined on other self-dual lattices.
The phase diagram of a self-dual hierarchical lattice has been numerically investigated by Nobre.\cite{Nobre}
According to his result, the multicritical point is located near the conjectured value, $p_c = 0.8902(4)$.

In addition, the conjecture also works on mutually dual pairs of lattices.\cite{TSN}
In this case, we obtain the relationship between the locations $p_1$ and $p_2$ of the multicritical points for the mutually dual pairs as follows
\begin{equation}
H(p_1) + H(p_2) = 1.\label{conjecture relation}
\end{equation}
This relation is supported by a consistent result within its numerical error bar for the $\pm J$ Ising model on the triangular and hexagonal lattices as $H(p_1) + H(p_2) = 1.002(3)$,\cite{Qeiroz} where $p_1$ and $p_2$ denote the location of the multicritical point on the triangular $p_1 = 0.835806$ and the hexagonal $p_2 = 0.932704$ lattices, respectively.\cite{NO}

However there are cases in which the relation (\ref{conjecture relation}) and the numerical results for three mutually dual pairs of hierarchical lattices show derivations by large amounts close to $2\%, $ $H(p_1)+H(p_2)=1.0172,0.9829,0.9911$.\cite{HB}
The technique by Hinczewski and Berker in Ref. [12] is based on an exact calculation through the renormalization group analysis on hierarchical lattices.

These results motivated us to study other hierarchical lattices to see if the conjecture gives exact solutions.
If it does not, the next question is why the prediction of the conjecture falls very close to numerical estimates in all cases.
To verify these points, we evaluated the relation (\ref{conjecture relation}) and its Gaussian version for other five mutually dual pairs of hierarchical lattices.
In addition, we also reexamined Eq. (\ref{conjecture}) for the $\pm J$ Ising model and Eq. (\ref{Gconjecture}) for the Gaussian model on several self-dual hierarchical lattices including the case investigated by Nobre.
\section{Hierarchical lattice}
In this section, we introduce the hierarchical lattice and renormalization group on it.\cite{BO,GK,KG}
The renormalization group analysis on the hierarchical lattice is an exact technique to obtain the location of the transition point, though it is difficult to obtain such an exact solution on regular lattices.

\begin{figure*}
\begin{center}
\scalebox{.6}{\includegraphics*[45mm,210mm][110mm,255mm]{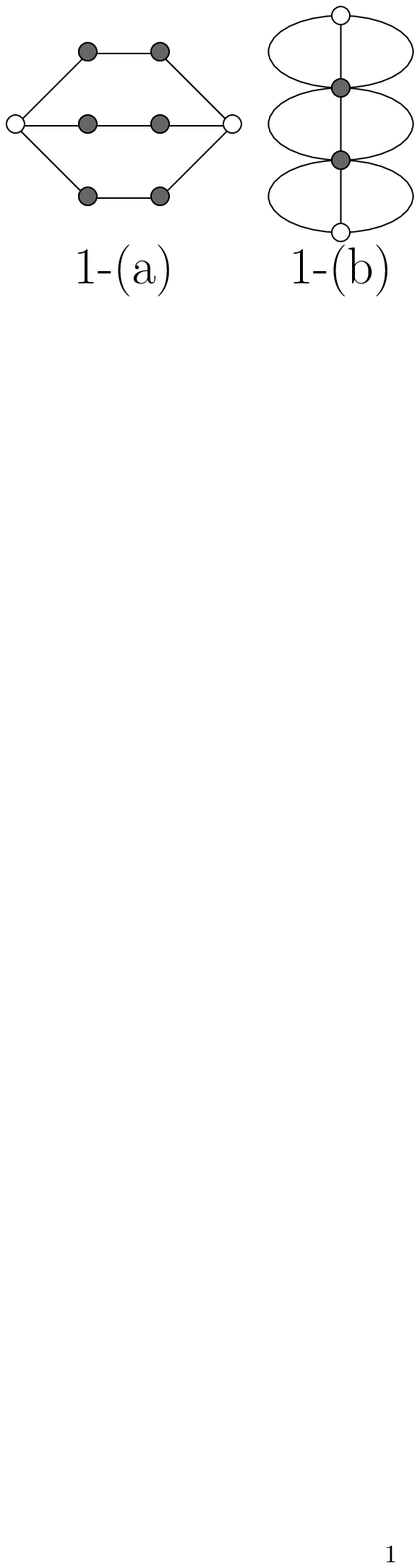}}
\scalebox{.6}{\includegraphics*[45mm,210mm][110mm,255mm]{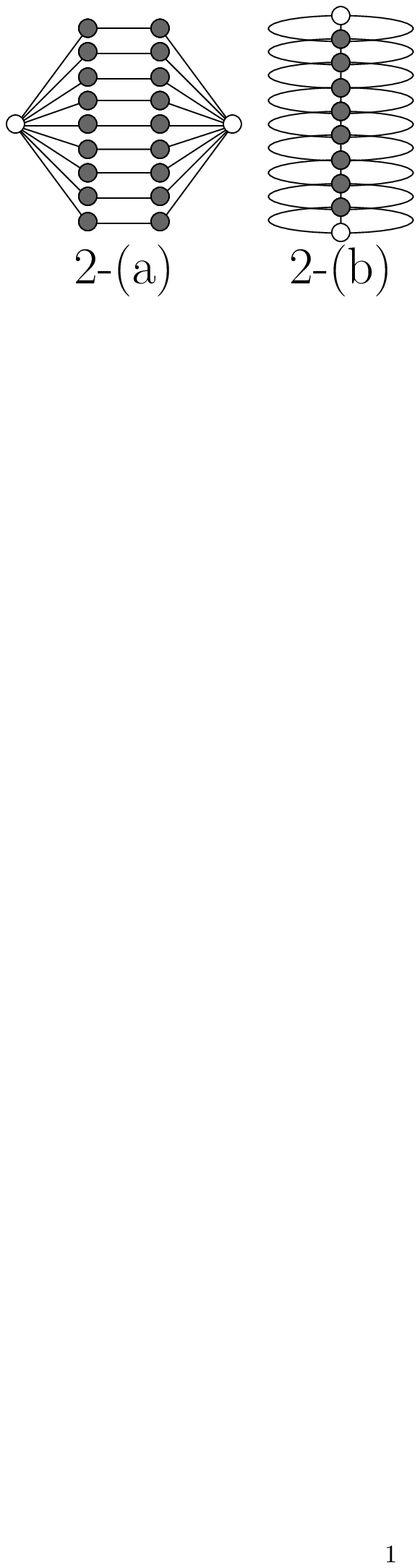}}
\scalebox{.6}{\includegraphics*[45mm,210mm][110mm,255mm]{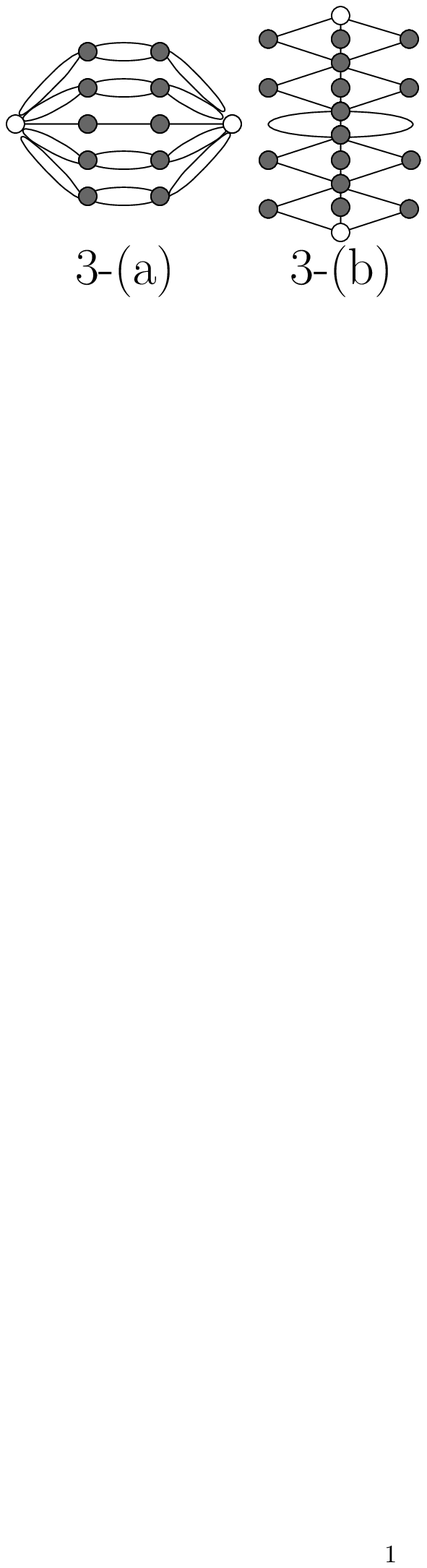}}
\end{center}
\caption{{\small 
Three mutually dual pairs of hierarchical lattices investigated in Ref. [12]. 
The numbers $1$, $2$, and $3$ denote these three mutually dual pairs of hierarchical lattices studied in the presented paper.}}\label{hierarchical1}
\end{figure*}

\begin{figure*}
\begin{center}
\scalebox{.52}{\includegraphics*[45mm,210mm][110mm,255mm]{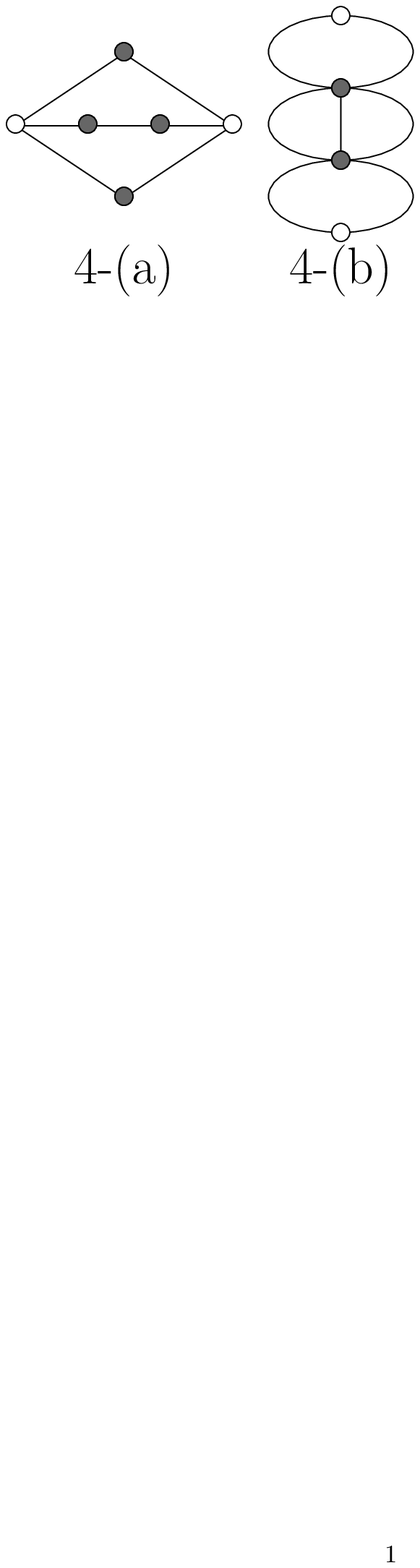}}
\scalebox{.52}{\includegraphics*[45mm,210mm][110mm,255mm]{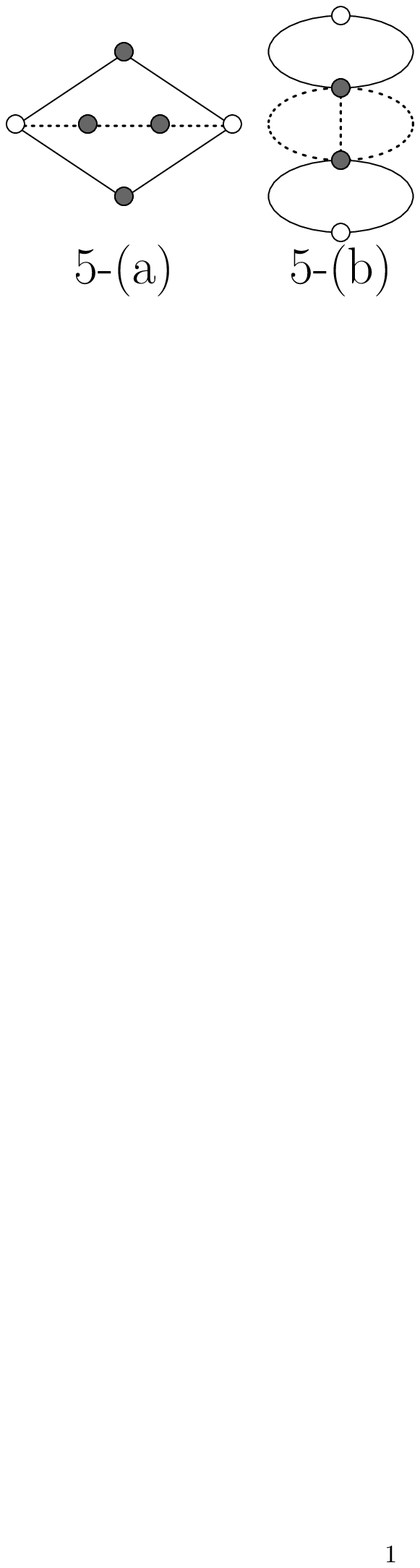}}
\scalebox{.52}{\includegraphics*[45mm,210mm][110mm,255mm]{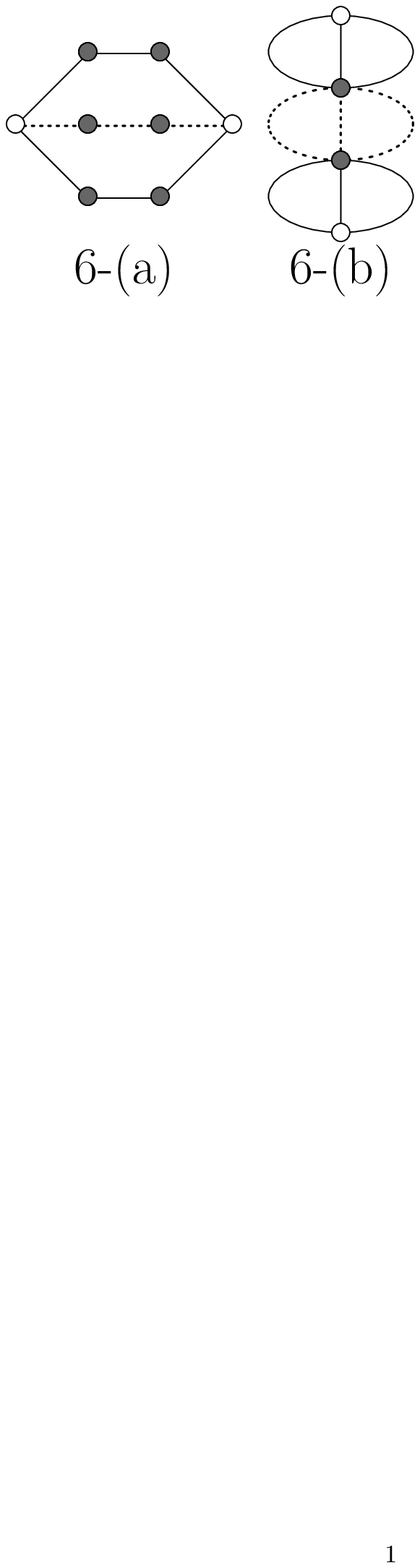}}
\scalebox{.52}{\includegraphics*[45mm,210mm][110mm,255mm]{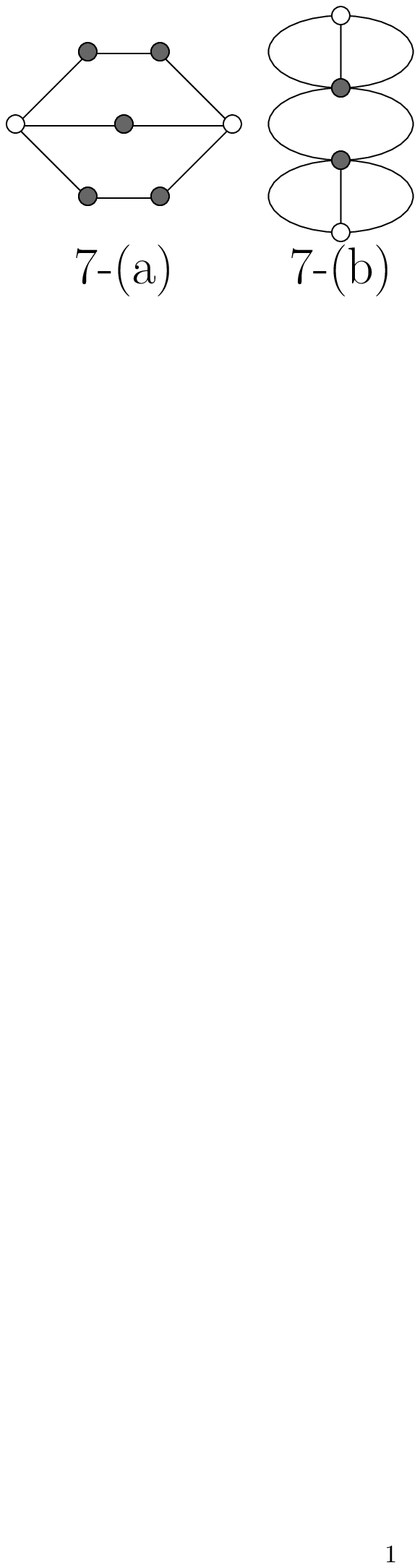}}
\scalebox{.52}{\includegraphics*[45mm,210mm][110mm,255mm]{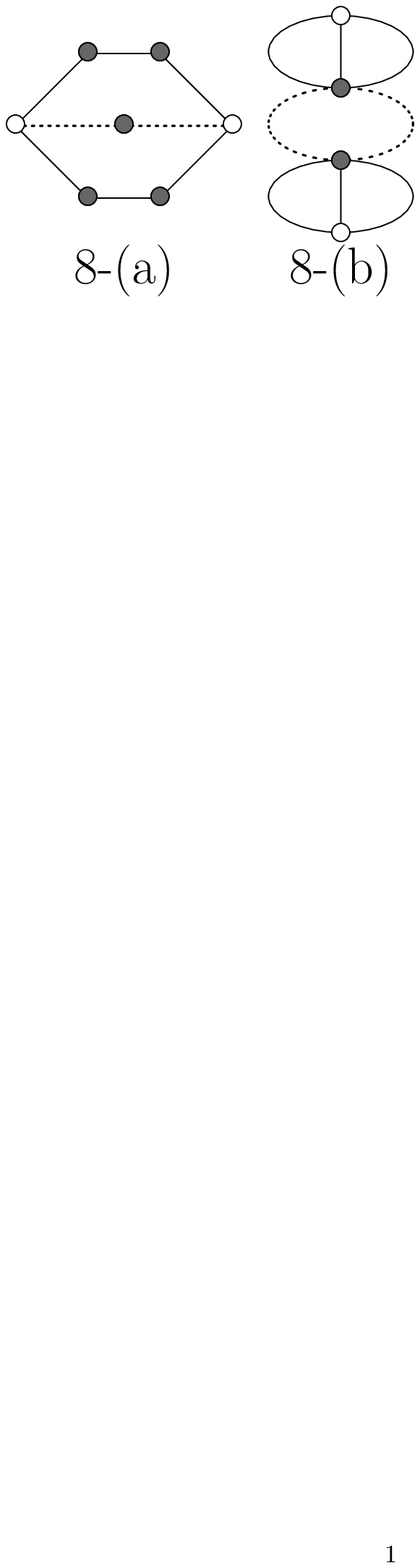}}
\end{center}
\caption{{\small 
Additional mutually dual pairs of hierarchical lattices studied in the presented paper. 
The solid lines denote bonds replaced by the renormalized bonds at each renormalization. Bonds shown dashed stay unrenormalized.}}\label{hierarchical2}
\end{figure*}

\begin{figure}
\begin{center}
\scalebox{.55}{\includegraphics{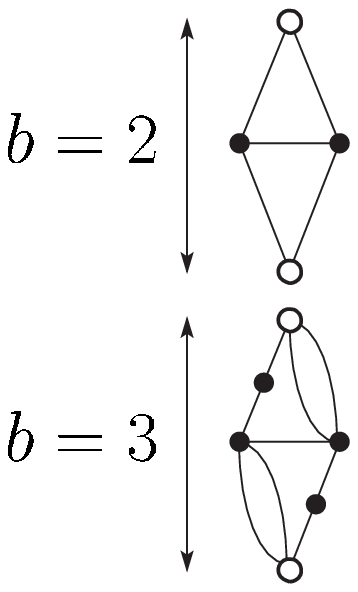}}
\scalebox{.55}{\includegraphics{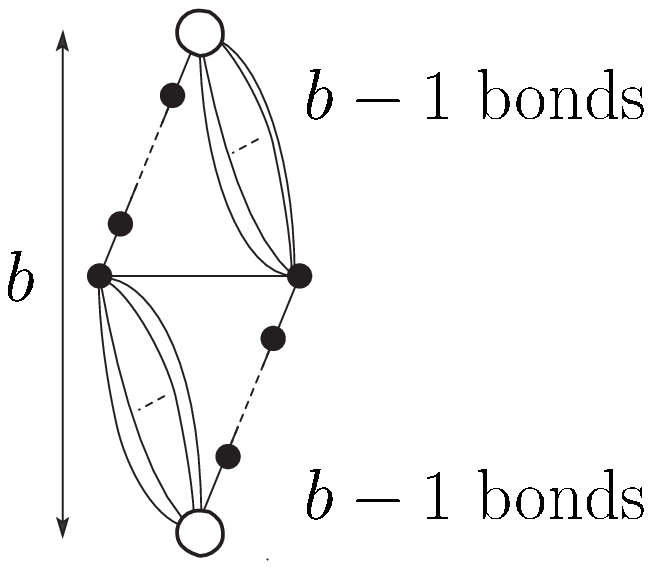}}
\scalebox{0.90}{\includegraphics{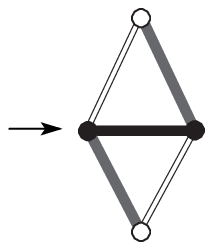}}
\end{center}
\caption{{\small 
Self-dual hierarchical lattices. 
After renormalization of bond moving and decimation, the self-dual hierarchical lattices become the structure like the Wheatstone bridge.
}}\label{self-dual}
\end{figure}

In the present paper, we investigate phase transitions for three mutually dual pairs of hierarchical lattices shown in Fig. \ref{hierarchical1} studied by Hinczewski and Berker\cite{HB} and additional five pairs shown in Fig. \ref{hierarchical2}.
We also examine phase transitions in several self-dual hierarchical lattices in Fig. \ref{self-dual}.
The scale factor $b$ denotes the length of the unit of the hierarchical lattice.
We examine phase transitions and estimate the location of the multicritical point, for $b=2$, $3$, $4$, $5$, and $6$ on the self-dual hierarchical lattices.

\begin{figure}
\begin{center}
\scalebox{0.90}{\includegraphics{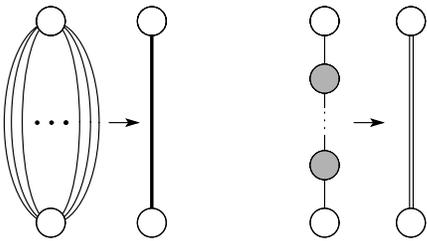}}
\end{center}
\caption{{\small Two renormalization steps.
The left-hand side is bond moving.
The right-hand side is decimation.}}\label{BMD}
\end{figure}

The renormalization group on these hierarchical lattices consists of two basic steps, which are known as the bond moving and decimation as in Fig. \ref{BMD}.
The black bold bond denotes the renormalized bond after bond moving, and the white bold bond expresses the renormalized bond after decimation.
\begin{figure}
\begin{center}
\includegraphics{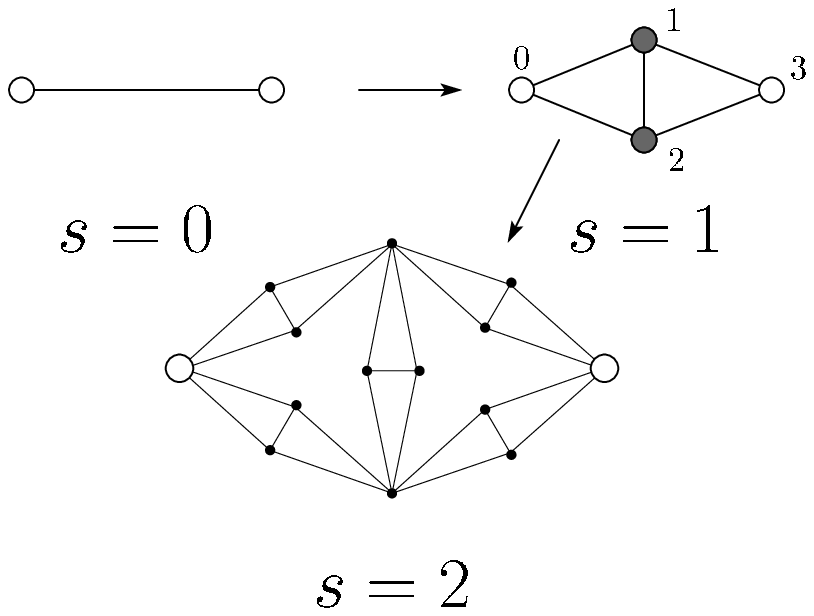}
\end{center}
\caption{{\small Construction of one of the self-dual hierarchical lattices. The number $s$ denotes the construction step.}}\label{hierarchical3}
\end{figure}
Construction of a hierarchical lattice starts from a single bond, and we iterate the process to substitute the single bond with a unit cell of more complex structure as in Fig. \ref{hierarchical3}.

Because a hierarchical lattice has an iterative structure consisting of unit cells as shown in Figs. \ref{hierarchical1}, \ref{hierarchical2} and \ref{self-dual}, we again obtain the same structure after we trace out degrees of freedom on each unit cell in renormalization group calculations, which are the inverse processes of the construction.
Therefore our task is to evaluate recursion relations of couplings on bonds, which relate sets of the couplings $\{K^{(r)}_{i'j'}\}$ after renormalization with $\{K^{(r-1)}_{ij}\}$ before renormalizations.
The superscript of $K^{(r)}_{ij}$ denotes the step of renormalization.
For example, the explicit recursion relations for the $b=2$ self-dual hierarchical lattice in Fig. \ref{hierarchical3}, are 
\begin{widetext}
\begin{eqnarray}
x^{(r)}_0 &=& \sum_{\{\sigma_i\}} \exp
\left( K^{(r-1)}_{01}\sigma_1 + K^{(r-1)}_{02}\sigma_2 +K^{(r-1)}_{12}\sigma_{1}\sigma_{2} + K^{(r-1)}_{13}\sigma_{1} + K^{(r-1)}_{23}\sigma_{2} \right) \label{recur1}\\
x^{(r)}_0{\rm e}^{-2K_{03}^{(r)}} &=& \sum_{\{\sigma_i\}} \exp
\left( K^{(r-1)}_{01}\sigma_1 + K^{(r-1)}_{02}\sigma_2 +K^{(r-1)}_{12}\sigma_{1}\sigma_{2} - K^{(r-1)}_{13}\sigma_{1} - K^{(r-1)}_{23}\sigma_{2} \right).\label{recur2}
\end{eqnarray}
\end{widetext}
Here $x_0^{(r)}$, which we call the principal Boltzmann factor, expresses the local (bond) Boltzmann factor for parallel spins on the ends of renormalized bonds.
The recursion relations (\ref{recur1}) and (\ref{recur2}) yield the renormalized principal Boltzmann factor and that for antiparallel spins on the ends of renormalized bonds, respectively.
The summation in the exponent is over all interactions in the unit cell.
The indices of couplings express bonds as labeled in Fig. \ref{hierarchical3}.

The partition function $Z_s$ for a hierarchical lattice after $s$-step construction is generally evaluated as
\begin{eqnarray}\nonumber
Z_s(\{K^{(0)}_{ij}\}) &\equiv& {x^{(0)}_0}^{N^{(s)}_B}z_{s}(\{K^{(0)}_{ij}\}) \\ \nonumber
&=& {x^{(1)}_0}^{N^{(s-1)}_B}z_{s-1}(\{K^{(1)}_{ij}\}) \\ \nonumber
&=& {x_0^{(2)}}^{N^{(s-2)}_B}z_{s-2}(\{K^{(2)}_{ij}\})\\
&=& \cdots = {x_0^{(s)}}^{N^{(0)}_B}z_{0}(\{K^{(s)}_{ij}\}),
\end{eqnarray}
where $N^{(s)}_B$ represents the number of bonds at the $s$th step of construction and $z_{s-r}$ is the partition function after $r$-step renormalization, which is normalized by ${x^{(r)}_0}^{N_B^{(s-r)}}$, namely $z_{s-r}\equiv \left(x_0^{(r)}\right)^{-N_B^{(s-r)}}Z_{s}$.
Because of this normalization, the value of $z_{s-r}$ is simply $2^{N_{s-r}}$ in the high-temperature limit and becomes $2$ in the low-temperature limit.\cite{Nstat}
Here $N_{s-r}$ denotes the number of sites after $r$ steps of renormalization for $s$ steps of construction.
In addition, one notices that the number of the construction step $s$ decreases effectively at each step of renormalization.

We obtain the free energy per site as,
\begin{eqnarray}\nonumber
& & -\beta f_s(\{K^{(0)}_{ij}\}) \\ \nonumber
& & = \frac{N^{(0)}_B}{N_s} \log x_0^{(s)}(\{K^{(s)}_{ij}\}) + \frac{1}{N_s}\log z_{0}(\{K^{(s)}_{ij}\}),\\
\end{eqnarray}
where $N_B^{(0)} = 1$.
Therefore the free energy per site of a model on a hierarchical lattice is generally written as, in the thermodynamic limit $s \to \infty$, 
\begin{eqnarray}\nonumber
& & \lim_{s\to \infty}-\beta f_{s}(\{K^{(0)}_{ij}\}) \\ \nonumber
& & = \lim_{s\to \infty}\left\{ \frac{1}{N_s} \log {x_0^{(s)}}(\{K^{(s)}_{ij}\}) + \frac{1}{N_s}\log z_{0}(\{K^{(s)}_{ij}\})\right\}. \\
\label{freeinf}
\end{eqnarray}
The last term in this equation can be calculated for the periodic and free boundary conditions by the fact that the hierarchical lattice becomes a single bond after sufficient steps of the renormalization as
\begin{equation}
z_{0}(\{K^{(\infty)}_{ij}\}) = 
\begin{cases}
 2 & {\rm (periodic)} \\
 2\left\{1+\exp(-2K^{(\infty)}_{ij}))\right\} & {\rm (free)}.
\end{cases}
\end{equation}
This is negligible due to $N_{\infty} \to \infty$ for the case of the periodic boundary condition.
This choice of the boundary condition does not affect the results.
Therefore only the first term $\log x_0^{(\infty)}$ in Eq. (\ref{freeinf}) is significant.
This quantity is a function of of renormalized couplings $\{K^{(\infty)}_{ij}\}$, which in general obeys a non-trivial distribution in quenched random systems.
In the next section, we observe the flow of the renormalized couplings $\{K^{(\infty)}_{ij}\}$ using a stochastic technique by Nobre\cite{Nobre} to investigate phase transitions on the hierarchical lattices, and estimate the location of the multicritical point for the $\pm J$ Ising model and the Gaussian model.
\section{Quenched systems}
In Nobre's implementation of renormalization group for disordered systems on hierarchical lattices,\cite{Nobre} we first produce a sample pool of interactions, following the initial distribution.
For example, our analysis starts from preparation of a pool following the $\pm J$ or Gaussian distribution.
Then we randomly choose bonds from this sampling pool and form a unit cell of the hierarchical lattice under consideration.
In this unit cell, we carry out the renormalization calculation using adequate equations such as Eqs. (\ref{recur1}) and (\ref{recur2}) and obtain renormalized interactions.
Iterating this procedure for the other bonds, we obtain another pool consisting of the renormalized interactions, which follows a new type of distribution function of the renormalized interactions.
Using this renormalized distribution, we reproduce a pool of the renormalized interactions, and iterate the above procedures while observing the moments of interactions $\langle K_{ij} \rangle$ and $\langle K^2_{ij} \rangle$ at every step where $\langle \cdots \rangle$ means the configurational average over the renormalized distribution.
If $\langle K_{ij} \rangle$ goes to infinity, it is considered that the renormalization flow of the sampling pool is attracted toward the ferromagnetic fixed point in the interaction space.
On the other hand, when $\langle K_{ij} \rangle$ goes to zero, two possible scenarios are considered.
To distinguish two different scenarios, we have to observe $\langle K^2_{ij} \rangle$.
If this moment goes to infinity, it is a signal that the sampling pool is attracted toward the spin glass fixed point.
Otherwise, we find a signal that $\langle K^2_{ij} \rangle$ falls zero, then the sampling pool goes to the paramagnetic fixed point.
An additional scenario is seen in the present study, which has not been investigated by Nobre's method yet.
As shown in Fig. \ref{hierarchical2}, the hierarchical lattices of type $5$, $6$ and $8$ include a part of interactions following the initial distribution function in the unit cell (see the dashed lines).
These interactions induce the possibility of a fixed line like the Kosterlitz-Thouless (KT) phase that the sampling pool goes toward neither $\langle K_{ij} \rangle \to \infty$ nor $\langle K_{ij} \rangle \to 0$.\cite{HB2}
This fixed line can be detected by $|K^{(n)}_{ij}-K^{(n-1)}_{ij}|\to 0$.

\begin{figure}[tb]
\begin{center}
\includegraphics[scale=0.6]{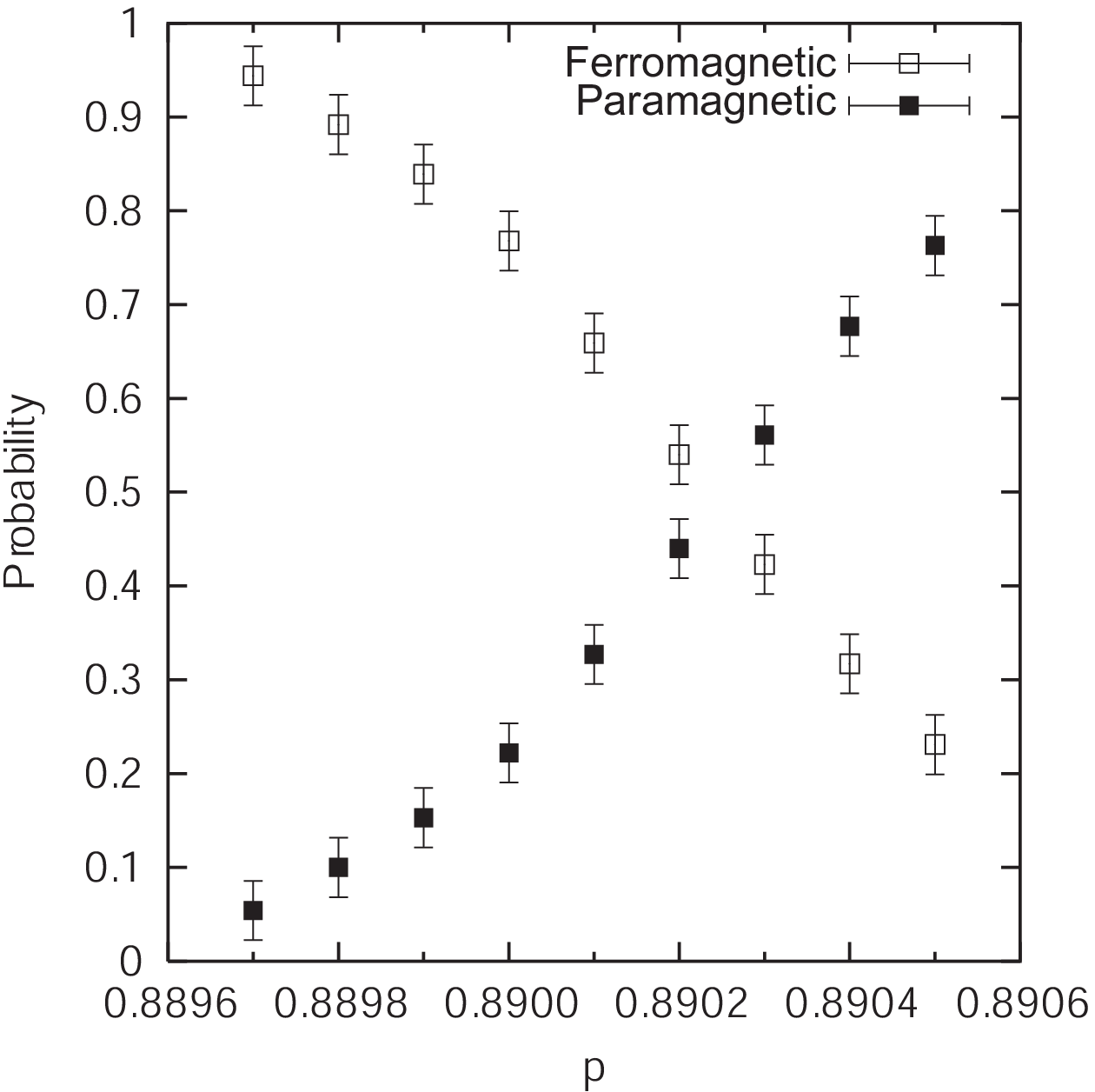}
\end{center}
\begin{center}
\includegraphics[scale=0.6]{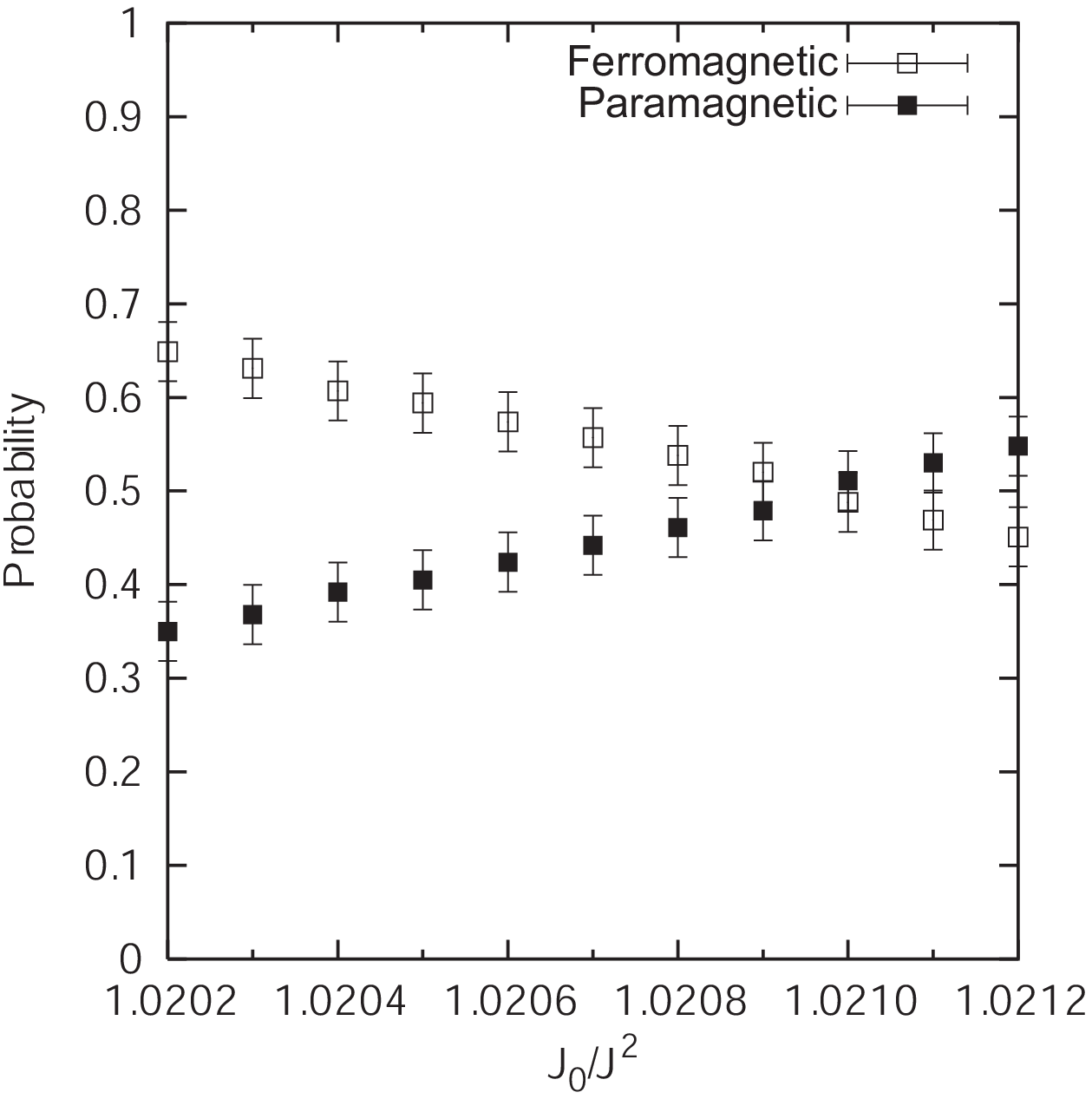}  
\end{center}  
\caption{{\small Results for the $\pm J$ (upper panel) and Gaussian (lower panel) Ising models along the Nishimori line on the self-dual hierarchical lattice with $b=3$, see Fig. \ref{self-dual}. 
The conjecture states that the multicritical points are located at $p_c = 0.889972$ and $J_0/J^2 = 1.02127$, respectively. 
The white square denotes the probability of the paramagnetic phase and the black one represents that of the ferromagnetic phase.
The error bars represent $1/\sqrt{1000}$ reflecting the number of the sampling pools.}
}\label{Nobreplot}
\end{figure}

We introduced $2,000,000$ bonds as the set of a sampling pool, and prepared $1,000$ sampling pools in the present study, except for the lattice number $5$ (a) and (b) in Fig. \ref{hierarchical2}, whose sampling pool has $1,800,000$ bonds.
We observed the resulting phases after $30$ steps of renormalization iterations.
For the hierarchical lattices with the possibility of a KT transition, we carry out the renormalization of $50$ more steps than the other hierarchical lattices.
Because the investigations are carried out for the hierarchical lattices of finite size and with finite number of bonds in the sampling pool, we cannot find clear boundary expressing the phase transition.
In fact, for a given lattice, some sampling pools go to the ferromagnetic fixed point (or KT phase) and others are attracted toward the paramagnetic fixed point (or KT phase).
We obtained the probabilities of appearance of each phase as a result by this method.
For example, the results for the $\pm J$ Ising model and Gaussian model on Nobre's self-dual hierarchical lattice depicted in Fig. \ref{self-dual} are shown in Fig. \ref{Nobreplot}.
For the other hierarchical lattices, we obtain similar results to these plots.
We explain the obtained plot in Fig. \ref{Size} below, which concerns the error bars for these investigations. 
In the thermodynamic limit, all the plots as in Fig. \ref{Nobreplot} become step functions.
However, we investigated the finite-size hierarchical lattices.
The slopes of all the plots are finite.
We have checked these finite-size effects as shown in Fig. \ref{Size}. 
From these analyses, the final error bars have been chosen to be $p_c/\sqrt{N_B}$ for $\pm J$ Ising model and $J_0/J^2\sqrt{N_B}$ for the Gaussian model, where $N_B$ is the number of bonds of the hierarchical lattice.

\begin{figure}
\begin{center}
\includegraphics[scale=0.6]{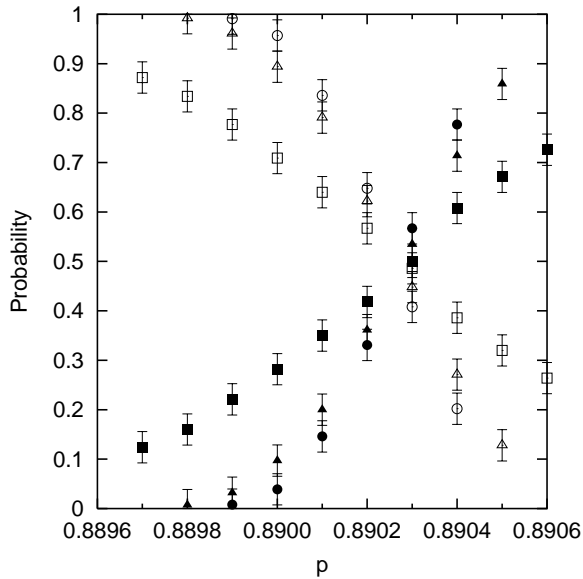}
\end{center}
\caption{{\small 
Size effect for the $\pm J$ Ising model on the self-dual hierarchical lattice with $b=3$. 
The black and white marks denote the probabilities of the paramagnetic and ferromagnetic phases, respectively.
The symbols {\footnotesize $\square$} and {\footnotesize $\blacksquare$} are for $10^3$ sampling pools with $10^6$ bonds, $\vartriangle$ and $\blacktriangle$ are for those with $5\times 10^6$ bonds, and {\large $\circ$} and {\large $\bullet$} is with $9 \times 10^6$ bonds.
}}
\label{Size}
\end{figure}
The results for the $\pm J$ model are given in Tables \ref{Results1} and \ref{Results2}, and those for the Gaussian model are in Table \ref{GResults}.
We show the values of the binary entropy $H(p)$ for comparison with the conjecture.
For the Gaussian Ising model, we also give the values of $H_G(J_0/J^2)$ for the self-dual hierarchical lattices.
Similarly to the case of the $\pm J$ Ising model, it can be shown that the summation of both of values of $H_G(J_0/J^2(A))$ and $H_G(J_0/J^2(B))$ should be unity for the mutually dual pairs.
We show such values for comparison in Table \ref{GResults}.
We express here pairs of the locations of the multicritical points as $J_0/J^2(A)$ and $J_0/J^2(B)$.
Seeing these results, we confirm slight but non-negligible deviations from unity for both cases of the $\pm J$ and the Gaussian models.
We find the general tendency that the difference from unity for the Gaussian Ising model is smaller than in the $\pm J$ Ising model.
\begin{table}
\begin{tabular}{ccc}
\hline
Lattice & $p_c$ & $2H(p_c)$  \\
\hline
$b=2$ self-dual & $0.8915(6)$ & $0.991(4)$ \\
$b=3$ self-dual & $0.8903(2)$ & $0.998(1)$ \\
$b=4$ self-dual & $0.8892(6)$ & $1.005(4)$ \\
$b=5$ self-dual & $0.8895(6)$ & $1.003(4)$ \\
$b=6$ self-dual & $0.8890(6)$ & $1.006(4)$ \\
$b=7$ self-dual & $0.8891(6)$ & $1.005(4)$ \\
$b=8$ self-dual & $0.8889(6)$ & $1.006(4)$ \\
\hline
\end{tabular}
\caption{{\small The locations of the multicritical points for the $\pm J$ Ising model on the self-dual hierarchical lattices.
Also shown are the values $2H(p_c)$, which should be unity according to the conjecture.}
}\label{Results1}
\end{table}

\begin{table}
\begin{tabular}{cccc}
\hline
Lattice & $p_1$ & $p_2$ & $H(p_1)+H(p_2)$  \\
\hline
1 & $0.9338(7)$ & $0.8265(6)$ & $1.017(4)$  \\
2 & $0.8149(6)$ & $0.9487(7)$ & $0.983(4)$  \\
3 & $0.7526(5)$ & $0.9720(7)$ & $0.991(5)$  \\
4 & $0.8712(6)$ & $0.9079(6)$ & $0.998(4)$  \\
5 & $0.8700(6)$ & $0.9081(7)$ & $1.000(4)$ \\
6 & $0.9337(7)$ & $0.8266(6)$ & $1.017(4)$ \\
7 & $0.9084(6)$ & $0.8678(6)$ & $1.005(4)$ \\
8 & $0.9065(6)$ & $0.8686(6)$ & $1.009(4)$ \\
\hline
\end{tabular}
\caption{{\small The locations of the multicritical points for the $\pm J$ Ising model on mutually dual pairs of the hierarchical lattices.
The results for lattices number 1 to 3 reproduce the results by Hinczewski and Berker.}
}\label{Results2}
\end{table}
\begin{table}
\begin{tabular}{ccc}
\hline
Lattice & $J_0$ & $2H_G(J_0)$  \\
\hline
$b=3$ self-dual & $1.0209(3)$ &  $1.0011(4)$  \\
\hline
\end{tabular}
\begin{tabular}{cccc}
\hline
Lattice & $J_0(A)$ & $J_0(B)$ & $H_G(J_0(A))+H_G(J_0(B))$ \\
\hline
1 & $0.7605(5)$ & $1.3174(9)$ & $1.0005(8)$  \\
2 & $0.7655(5)$ & $1.3118(9)$ & $1.0000(8)$  \\
3 & $0.5569(4)$ & $1.6151(11)$ & $0.9999(7)$ \\
4 & $0.9704(7)$ & $1.0730(8)$ & $1.0009(10)$  \\
5 & $0.9701(7)$ & $1.0733(8)$ & $1.0009(10)$  \\
6 & $1.3175(9)$ & $0.7606(5)$ & $1.0003(8)$  \\
7 & $1.1450(8)$ & $0.9040(6)$ & $1.0001(9)$  \\
8 & $1.1436(8)$ & $0.9055(6)$ & $1.0005(9)$  \\
\hline
\end{tabular}
\caption{{\small The locations of the multicritical points for the Gaussian model on the $b=3$ self-dual and mutually dual pairs of the hierarchical lattices.}
}\label{GResults}
\end{table}

\section{Replicated systems}
There are slight differences between the results by the conjecture and the numerical data by the renormalization group analysis for the multicritical points of quenched systems as shown in the previous section.
We examine the conjecture for replicated systems on the self-dual hierarchical lattices in this section.
Because the conjecture is based on the duality and the replica method,\cite{NN,MNN,TSN,NO,Nstat} it is expected that we find such discrepancies also for replicated systems with the replica number $n$ of natural numbers as in the quenched system ($n \to 0$).

If the partition function is a single-variable function, we can obtain the transition point as the fixed point of the duality.
Then equation $x_0(K)=x_0^*(K)$ gives the exact transition point, where $x_0(K)$ and $x_0^*(K)$ are the original and dual principal Boltzmann factors.\cite{NN,MNN,Nstat,TSN,NO}
We illustrate this point by the pure Ising model as
\begin{eqnarray}
x_0(K) &=& {\rm e}^{K} \\
x^*_0(K) &=& \frac{{\rm e}^K+ {\rm e}^{-K}}{\sqrt{2}}.
\end{eqnarray}
By equating $x_0(K)$ and $x_0^*(K)$, we find the transition point ${\rm e}^{-2K_c} = \sqrt{2}-1$ for the pure Ising model on the self-dual square lattice. 
We assume that the equation $x_0(K)=x_0^*(K)$ is also satisfied at the multicritical point for the replicated systems as well as for the quenched system ($n \to 0$), though the replicated systems have complicated interactions.\cite{NN,MNN,Nstat,TSN,NO}

Let us set $K=K_p$, which is the condition of the Nishimori line, for the replicated $\pm J$ Ising model.
The quantity $K_p$ is defined as ${\rm e}^{-2K_p} = (1-p)/p$.
Both of the principal Boltzmann factors are then given as,\cite{NN,MNN}
\begin{eqnarray}
x_0(K) &=& 2\cosh\{ (n+1)K \} \\
x^*_0(K) &=& 2^{\frac{n}{2}} \cosh^nK.
\end{eqnarray}
Equation $x_0(K) = x_0^*(K)$ gives the conjecture (\ref{conjecture}) in the limit $n \to 0$.
Validity of this conjecture can be rigorously shown for $n=1$ and $2$, and for $n=3$, the same has been numerically confirmed for the square lattice.\cite{MNN}
It is worthwhile to examine whether $x_0(K) = x^*_0(K)$ is satisfied or not at the multicritical point for the replicated $\pm J$ Ising model on the self-dual hierarchical lattices.
Because the replicated $\pm J$ Ising model does not have any randomness for couplings, which has been taken into consideration by the configurational average, we can derive directly the multicritical points, by evaluating recursion relations such as in Eqs. (\ref{recur1}) and (\ref{recur2}).

We obtained the locations of the multicritical points on the several self-dual hierarchical lattices from $b=2$ to $b=6$ with the replica number $n=1$, $2$, $3$, and $4$.
The results are shown and compared with the predictions by the conjecture $x_0(K) = x^*_0(K)$ in Table \ref{replica}.
\begin{table}
\begin{tabular}{ccccc}
\hline
$b$ & $n$ & $p_c$ & $p_{\rm numerical}$ & $p_c-p_{\rm numerical}$\\
\hline

$2$   & $n\to0$ & $ 0.889972$ & $0.8915(6)$ & $-0.0015(6)$ \\
      & $1$ & $ 0.821797$ & $0.821797$ & $0$ \\
      & $2$ & $ 0.788675$ & $0.788675$ & $0$ \\
      & $3$ & $ 0.769563$ & $0.768851$ & $0.000713$ \\
      & $4$ & $ 0.757348$ & $0.755451$ & $0.001897$ \\
\hline
$3$   & $n\to0$ & $ 0.889972$ & $0.8903(2)$ & $-0.0003(2)$ \\
      & $1$ & $ 0.821797$ & $0.821797$ & $0$ \\
      & $2$ & $ 0.788675$ & $0.788675$ & $0$ \\
      & $3$ & $ 0.769563$ & $0.769022$ & $0.000542$ \\
      & $4$ & $ 0.757348$ & $0.755942$ & $0.001406$ \\
\hline
$4$   & $n\to0$ & $ 0.889972$ & $0.8892(6)$ & $0.0007(6)$ \\
      & $1$ & $ 0.821797$ & $0.821797$ & $0$ \\
      & $2$ & $ 0.788675$ & $0.788675$ & $0$ \\
      & $3$ & $ 0.769563$ & $0.769649$ & $-0.000086$ \\
      & $4$ & $ 0.757348$ & $0.757763$ & $-0.000415$ \\
\hline
$5$   & $n\to0$ & $ 0.889972$ & $0.8895(6)$ & $0.0004(6)$ \\
      & $1$ & $ 0.821797$ & $0.821797$ & $0$ \\
      & $2$ & $ 0.788675$ & $0.788675$ & $0$ \\
      & $3$ & $ 0.769563$ & $0.7705020$ & $-0.000939$ \\
      & $4$ & $ 0.757348$ & $0.7601328$ & $-0.002785$ \\
\hline
$6$   & $n\to0$ & $ 0.889972$ & $0.8890(6)$ & $0.0010(6)$ \\
      & $1$ & $ 0.821797$ & $0.821797$ & $0$ \\
      & $2$ & $ 0.788675$ & $0.788675$ & $0$ \\
      & $3$ & $ 0.769563$ & $0.771376$ & $-0.001813$ \\
      & $4$ & $ 0.757348$ & $0.762313$ & $-0.004965$ \\
\hline
\end{tabular}
\caption{{\small Differences between $p_c$ by the conjecture equation $x_0(K) = x_0^*(K)$, and $p_{\rm numerical}$ by the exact renormalization analysis for the $n$-replicated $\pm J$ Ising model on several self-dual hierarchical lattices.
For $n \to 0$, $p_{\rm numerical}$ denotes the results obtained by Nobre's technique.}}
\label{replica}
\end{table}
The results obtained in the previous section for the quenched system ($n \to 0$) are also shown for comparison.
Equation $x_0(K) = x^*_0(K)$ gives the exact answer for $n=1$ and $2$.
For $n=3$ and $4$, the multicritical point locates slightly away from the results of the conjecture.
Therefore the assumption of the validity for the conjecture is violated for the self-dual hierarchical lattices.
Considering $b\to \infty$, we find that the conjecture does not always work well on this self-dual hierarchical lattice, since the system becomes an effectively one-dimensional chain without finite-temperature transition in the limit $b\to \infty$ although the conjecture gives the results independent of $b$.
\section{Improvement of the conjecture}
\begin{figure}
\begin{center}
\includegraphics{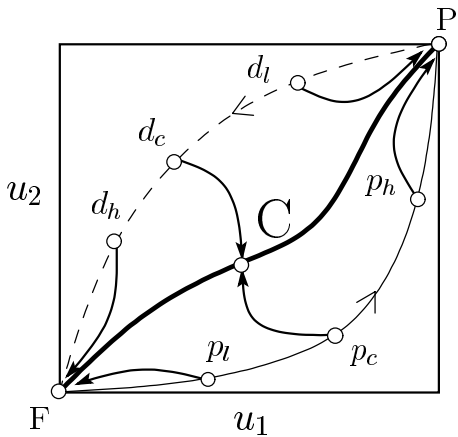}
\end{center}  
\caption{{\small 
A schematic picture to consider the renormalization flow and the duality for the replicated $\pm J$ Ising model.
}}
\label{Trajectory}
\end{figure}

In this section we propose an improvement of the conjecture, which reduces discrepancies observed in Table \ref{replica}.
We here consider the partition function for the replicated $\pm J$ Ising model and its dual one and discuss their relationship following Ref. [9].
The partition function for the replicated $\pm J$ Ising model on the Nishimori line is a multi-variable function of the Boltzmann factors as
\begin{equation}
Z(K) = {x_0(K)}^{N_B}z(u_1,u_2,\cdots,u_n),
\end{equation}
where the $u_r$ are the relative Boltzmann factors defined as
\begin{equation}
u_r(K) = \frac{x_r(K)}{x_0(K)} = \frac{\cosh\{(n+1-2r)K\}}{\cosh\{(n+1)K\}}.
\end{equation}
Here $r$ denotes the number of antiparallel pairs among the $n$ pairs.
The duality gives the following relationship between the original and dual partition functions,\cite{Nstat}
\begin{eqnarray}\nonumber
& & {x_0(K)}^{N_B}z(u_1,u_2,\cdots,u_n) \\
& & \quad = {x^*_0(K)}^{N_B}z(u^*_1,u^*_2,\cdots,u^*_n),\label{PF}
\end{eqnarray}
where the $u_r^*$ are the dual relative Boltzmann factors defined as
\begin{equation}
u^*_r(K) =
\begin{cases}
 \tanh^r K  & (r={\rm even})\\
 \tanh^{r+1} K & (r={\rm odd}).
\end{cases}
\end{equation}
Figure \ref{Trajectory} shows the relationship between the curves $(u_1(K),u_2(K)),\cdots,u_n(K))$ (the thin curve going through $p_c$) and $(u^*_1(K),u^*_2(K)),\cdots,u^*_n(K))$ (the dashed line).
For convenience, we show the projections on the two-dimensional plane $(u_1,u_2)$.
For $T \to 0$ $(K \to \infty)$, $u_r(K) \to 0$ for any $r$, corresponding to the point F in Fig. \ref{Trajectory}.
As $T$ changes from $0$ to $\infty$, the point representing $(u_1(K),u_2(K)),\cdots,u_n(K))$ moves toward the point P along the thin line in Fig. \ref{Trajectory}. 
Then the corresponding dual point $(u^*_1(K),u^*_2(K)),\cdots,u^*_n(K))$ moves along the dashed line in the opposite direction from P to F.

If we were to consider a model with a single-variable partition function, the thin curve would overlap the dashed line, which fact would be reflected in the relation $u^*_r(K) = u_r(K^*)$.
Solving this relation, we obtain the duality relation for the coupling constant $K^*(K)$.
For example, the pure Ising model has the reduced Boltzmann factors as $u_1(K) = {\rm e}^{-2K}$ and $u^*_1(K) = \tanh K$.
We obtain the duality relation ${\rm e}^{-2K^*} = \tanh K$ from $u^*_1(K) = u_1(K^*)$.
However the replicated $\pm J$ Ising model is given by the multi-variable partition function of $(u_1(K),u_2(K)),\cdots,u_n(K))$.
The thin curve $(u_1(K),u_2(K)),\cdots,u_n(K))$ does not coincide with the dashed curve $(u^*_1(K),u^*_2(K)),\cdots,u^*_n(K))$.
Therefore we cannot identify the critical point with the fixed point of duality.

A new point of view from renormalization group helps us proceed further.
Let us notice two facts concerning renormalization group transformation.
(i) The critical point is attracted toward the unstable fixed pint.
(ii) The partition function does not change its functional form by renormalization for hierarchical lattice; only the values of arguments change.
Therefore the renormalized system also has a representative point in the same space $(u_1(K),u_2(K)),\cdots,u_n(K))$ as in Fig. \ref{Trajectory}, with the renormalization flow following the arrows emanating from $p_c$ and $d_c$ to C, $p_h$ and $d_l$ to P, and $p_l$ and $d_h$ to F.
We express such a development of relative Boltzmann factors at each renormalization step on the $n$-dimensional hyperspace as $(u^{(r)}_1,u^{(r)}_2,\cdots,u^{(r)}_n)$, where the superscript means the number of renormalization steps. 
The renormalization flow from the critical point $p_c$ reaches the fixed point C, $(u^{(\infty)}_1,u^{(\infty)}_2,\cdots, u^{(\infty)}_n)$.
On the other hand, there is the point $d_c$ related with $p_c$ by the duality.
We expect that the renormalization flow from this dual point $d_c$ also reaches the same fixed point C because $p_c$ and $d_c$ represent the same critical point due to Eq. (\ref{PF}).

Considering the above property of the renormalization flow as well as the duality, we find that the duality relates two trajectories of the renormalization flow from $p_c$ and from $d_c$, tracing the renormalization flows at each renormalization.
In other words, after a sufficient number of renormalization steps, the thin curve representing the original system and the dashed curve for the dual system both approach the common renormalized system depicted as the bold line in Fig. \ref{Trajectory}, which goes through the fixed point C.

It is expected that the partition function therefore becomes a single-variable function described by the bold curve.
This fact enables us to find the duality relation and identify the multicritical point by the following equation,
\begin{equation}
x^{(\infty)}_0(K) = {x_0^*}^{(\infty)}(K),\label{imcon}
\end{equation}
similarly to $x_0(K)=x^*_0(K)$ for the pure Ising model.
We therefore have to evaluate Eq. (\ref{imcon}), not the relation $x_0(K)={x_0^*}(K)$ for the unrenormalized, bare quantities, to obtain the precise location of the multicritical point on the hierarchical lattices.
Equation (\ref{imcon}) is expected to predict the exact location of the multicritical point for the hierarchical lattice.

It should be noticed that the relation $x_0(K)={x_0^*}(K)$ predicts values very close to numerical estimates in many cases of regular lattices as indicated in Table \ref{Comparison}.
This means that the effects of renormalization are not large for those systems.
If we regard the relation $x_0(K)={x_0^*}(K)$ as the zeroth approximation for the location of the multicritical point, it is expected that the relation $x^{(1)}_0(K)={x_0^*}^{(1)} (K)$ is the first approximation and leads to more precise results than the relation $x_0(K)={x_0^*}(K)$ does.
We therefore propose the first-approximation equation $x^{(1)}_0={x_0^*}^{(1)}$ as the improved conjecture.
We evaluate the performance of this approximation in the next section for hierarchical lattices.
\section{Results by the improved conjecture}
In this section, we report the results by the improved conjecture $x^{(1)}_0(K)={x_0^*}^{(1)}(K)$ and evaluate its performance compared with the conventional conjecture.

The relation $x_0(K)={x_0^*}(K)$ of the conventional conjecture yields an equation that the binary entropy $H(p)$ equals to $1/2$ for self-dual hierarchical lattices as in Eq. (\ref{conjecture}).
Similarly to this relation, the improved conjecture $x^{(1)}_0(K)={x_0^*}^{(1)}(K)$ gives an equation in terms of the binary entropy given by the values of the renormalized couplings as described below.
After one-step renormalization, we obtain again the replicated Ising model on the hierarchical lattice with the renormalized couplings $\{K^{(1)}_{ij}\}$ and their distribution function $P^{(1)}(K_{ij})$.
Here the renormalized quantities are determined by the initial condition.
The original and dual principal Boltzmann factors for the replicated Ising model after one-step renormalization are given as
\begin{eqnarray}
x_0^{(1)}(K) &=& \int d K_{ij} P^{(1)}(K_{ij}) {\rm e}^{nK_{ij}} \\ \nonumber
x_0^{*(1)}(K) &=& \int d K_{ij} P^{(1)}(K_{ij}) \left(\frac{{\rm e}^{K_{ij}} + {\rm e}^{-K_{ij}}}{\sqrt{2}}\right)^n,\\
\end{eqnarray}
where the distribution function is given by, with the couplings $\{K^{(1)}_{ij}\}$ obtained by such Eqs. (\ref{recur1}) and (\ref{recur2}),
\begin{eqnarray}\nonumber
& & P^{(1)}(K_{ij}) \\ \nonumber
& & = \int \left\{ \prod_{{\rm unit}} d K^{(0)}_{ij} P(K^{(0)}_{ij}) \right\} \delta(K_{ij}-K^{(1)}_{ij}(\{K^{(0)}_{ij}\})).\\
\end{eqnarray}
The product runs over the bonds on the unit cluster of the hierarchical lattice.
Using these principal Boltzmann factors, we take the leading term of the replica number $n \to 0$ of the equation $x_0^{(1)}(K) = {x^*_0}^{(1)}(K)$ and obtain the improved conjecture for the quenched system as
\begin{equation}
\int d K_{ij} P^{(1)}(K_{ij}) \log_2 \left\{ 1 + \exp \left(-2K_{ij} \right) \right\} = \frac{1}{2}.\label{improvedcon}
\end{equation}
The left-hand side of this equation will be written as $H^{(1)}(p)$.
Equation (\ref{improvedcon}) gives the results for the replica number $n \to 0$ shown in Table \ref{improvedconjecture}.
Similarly, we can obtain an equation to predict the multicritical point for the replicated $\pm J$ Ising model with a finite replica number $n$, whose solutions for $n=1$, $2$, $3$, and $4$ also shown in Table \ref{improvedconjecture}.
All results are in excellent agreement with the numerical ones within their error bars.
Comparison of Table \ref{improvedconjecture} with the Table \ref{replica} clearly indicates significant improvements.
We also find the improved conjecture gives results depending on the feature of each hierarchical lattice because the prediction for the self-dual hierarchical lattice is different from each other, which was not the case before as seen in Table \ref{replica}.
\begin{table}
\begin{tabular}{ccccc}
\hline
$b$ & $n$ & $p_c$ & $p_{\rm numerical}$ & $p_c-p_{\rm numerical}$\\
\hline

$2$   & $0$ & $ 0.892025$ & $0.8915(6)$ & $-0.0005(6)$ \\
      & $1$ & $ 0.821797$ & $0.821797$ & $0$ \\
      & $2$ & $ 0.788675$ & $0.788675$ & $0$ \\
      & $3$ & $ 0.769048$ & $0.768851$ & $0.000197$ \\
      & $4$ & $ 0.755986$ & $0.755451$ & $0.000535$ \\
\hline
$3$   & $0$ & $ 0.890340$ & $0.8903(2)$ & $0.0000(2)$ \\
      & $1$ & $ 0.821797$ & $0.821797$ & $0$ \\
      & $2$ & $ 0.788675$ & $0.788675$ & $0$ \\
      & $3$ & $ 0.769138$ & $0.769022$ & $0.000116$ \\
      & $4$ & $ 0.756250$ & $0.755942$ & $0.000308$ \\
\hline
$4$   & $0$ & $ 0.889204$ & $0.8892(6)$ & $0.0000(6)$ \\
      & $1$ & $ 0.821797$ & $0.821797$ & $0$ \\
      & $2$ & $ 0.788675$ & $0.788675$ & $0$ \\
      & $3$ & $ 0.769629$ & $0.769649$ & $-0.000020$ \\
      & $4$ & $ 0.757619$ & $0.757763$ & $-0.000144$ \\
\hline
$5$   & $0$ & $ 0.889522$ & $0.8895(6)$ & $0.0000(6)$ \\
      & $1$ & $ 0.821797$ & $0.821797$ & $0$ \\
      & $2$ & $ 0.788675$ & $0.788675$ & $0$ \\
      & $3$ & $ 0.769968$ & $0.770502$ & $-0.000534$ \\
      & $4$ & $ 0.758461$ & $0.760133$ & $-0.001672$ \\
\hline
$6$   & $0$ & $ 0.889095$ & $0.8890(6)$ & $0.0000(6)$ \\
      & $1$ & $ 0.821797$ & $0.821797$ & $0$ \\
      & $2$ & $ 0.788675$ & $0.788675$ & $0$ \\
      & $3$ & $ 0.769947$ & $0.771376$ & $-0.001429$ \\
      & $4$ & $ 0.758300$ & $0.762313$ & $-0.004013$ \\
\hline
\end{tabular}
\caption{{\small The results by the improved conjecture $x^{(1)}_0(K)={x_0^*}^{(1)}(K)$.}}
\label{improvedconjecture}
\end{table}

\begin{figure}
\begin{center}
\includegraphics[scale=.9]{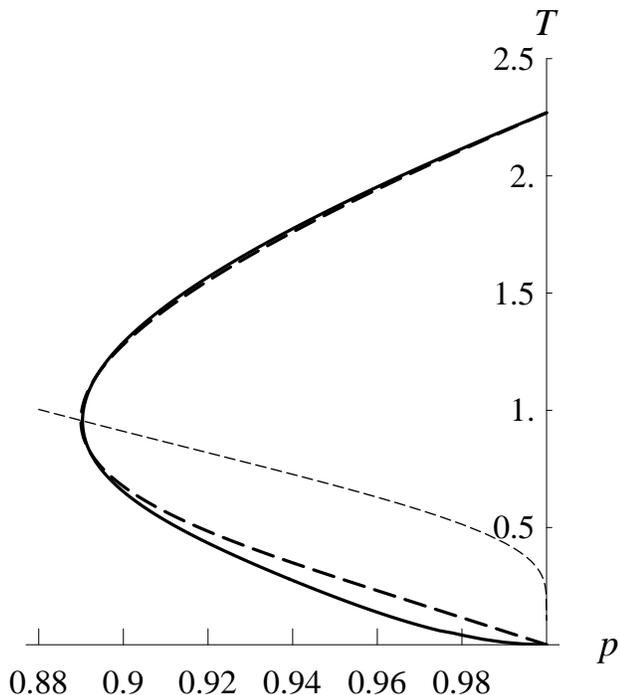}
\end{center}
\caption{{\small The phase boundary by the conventional and improved conjectures for a self-dual hierarchical lattice with $b=3$. The vertical axis is the temperature, and the horizontal axis is the probability for $J_{ij}=J>0$ of the $\pm J$ Ising model. The bold dashed line is by the conventional conjecture and the bold solid line is by the improved conjecture. The thin dashed line is the Nishimori line.
}}
\label{Phasediagram}
\end{figure}

We can also see the performance of the improved conjecture from another point of view.
We can predict the phase boundary by the conventional conjecture if we do not restrict ourselves to the Nishimori line $K_p = K$.
The well-known transition point $T_c = 2.26919$ for the pure Ising model is exactly reproduced by the conventional conjecture.
However, except for this transition point, the conventional conjecture fails to derive the precise phase boundary especially below the Nishimori line as seen in Fig. \ref{Phasediagram}, because the phase boundary of the $\pm J$ Ising model is expected to be vertical or slightly reentrant below the multicritical point.\cite{HN81,HNbook,Nobre,MB}
We find also inaccuracy of the conventional conjecture in the slope of the phase boundary at the transition point of the pure Ising model given as
\begin{equation}
\frac{1}{T_c}\left.\frac{dT}{dp} \right|_{T=T_c} = 3.43294.
\end{equation}
However this value is estimated as $3.23(3)$ by Nobre's method\cite{Nobre} on the $b=3$ self-dual hierarchical lattice and as $3.209$ by the exact perturbation for the square lattice.\cite{Domany}
The improved conjecture, on the other hand, yields for $b=3$
\begin{equation}
\frac{1}{T_c}\left.\frac{dT}{dp} \right|_{T=T_c} = 3.30712.
\end{equation}
This value is closer to $3.23(3)$.
Also the phase boundary below the Nishimori line is modified as moving toward the $p$-axis as in Fig. \ref{Phasediagram}.
Thus the improved conjecture works better than the conventional conjecture for describing the phase boundary.

The improved conjecture also succeeds in leading to the relation between the multicritical points on the mutually dual pairs.
It is straightforward to apply the improved conjecture to the mutually dual pairs, similarly to the case of the conventional conjecture\cite{TSN} as,
\begin{equation}
H^{(1)}(p_1) + H^{(1)}(p_2) = 1,\label{improvedrelation}
\end{equation}
where $p_1$ and $p_2$ denote the locations of the multicritical points on mutually dual pair.
We estimate the values of the left-hand side of Eq. (\ref{improvedrelation}) for several pairs of hierarchical lattices in Figs. \ref{hierarchical1} and \ref{hierarchical2}.
The estimated results are given in Table \ref{improvedmdp}.
We use the values of the locations of the multicritical points obtained by Nobre's method, as in Table \ref{Results2}, to compare the performance of the improved conjecture with that of the conventional conjecture $H(p_1)+H(p_2) = 1$.
\begin{table}
\begin{tabular}{cccc}
\hline
Lattice & $p_1$ & $p_2$ & value \\
\hline
1 & $0.9338(7)$ & $0.8265(6)$ & $1.002(7)$ \\
2 & $0.8149(6)$ & $0.9487(7)$ & $0.984(9)$ \\
3 & $0.7526(5)$ & $0.9720(7)$ & $0.993(9)$ \\
4 & $0.8712(6)$ & $0.9079(6)$ & $1.007(6)$ \\
5 & $0.8700(6)$ & $0.9081(7)$ & $1.011(6)$ \\
6 & $0.9337(7)$ & $0.8266(6)$ & $1.003(7)$ \\
7 & $0.9084(6)$ & $0.8678(6)$ & $0.996(6)$ \\
8 & $0.9065(6)$ & $0.8686(6)$ & $1.003(6)$ \\
\hline
\end{tabular}
\caption{{\small The results by the improved conjecture for the mutually dual pairs.
We estimate values of the left-hand side of Eq. (\ref{improvedrelation}) by the improved conjecture, shown on the right-most column of this Table. 
}
}\label{improvedmdp}
\end{table}
There are cases in which the improved conjecture agrees with the numerical estimates for the mutually dual pairs, for the lattices of type 1, 3, 6, 7, and 8 hierarchical lattices in Figs. \ref{hierarchical1} and \ref{hierarchical2}.
Unfortunately, we find three cases for the lattices of type 2, 4, and 5 in which the value of the left-hand side of the relation (\ref{improvedrelation}) is not unity within the error bars.
However we find impressive improvements in Table. \ref{improvedmdp} compared with the previous results in Table. \ref{Results2}.
\section{Discussions}
In the present paper, we first showed the existence of slight differences between the conventional conjecture and the numerical results for the locations of the multicritical points on several hierarchical lattices.
These discrepancies for the quenched system are caused by violation of satisfaction of the equation $x_0(K)=x_0^*(K)$ for the replicated systems.
This equation $x_0(K) = x_0^*(K)$ is satisfied for the case that the partition function is written by a single variable as in the pure Ising model.
We expect that the partition function can be written as a single-variable function after a sufficient number of renormalization steps, considering the fact that the plot describing the original model overlaps that of the dual model as in Fig. \ref{Trajectory}.
Based on this consideration, we proposed the improved conjecture as the first approximation of the exact relation to determine the critical point.
Through the derivation of the improved conjecture, one finds that the multicritical point on the self-dual lattice is given as a special point where the binary entropy given by the renormalized values on the unit of the hierarchical lattice becomes one half.
If we need the very precise location of the multicritical point, we may use the numerical methods for the renormalization group analysis to evaluate Eq. (\ref{imcon}).

The present study also gives a basis for the improvement of the conjecture on regular lattices.
The improved conjecture for the hierarchical lattice reflects individual characteristics of each hierarchical lattice, because it includes the renormalized couplings and corresponding distribution function, which depend on the structure of the hierarchical lattice under consideration.
Similarly to the case of such hierarchical lattices, if we adequately carry out the renormalization for regular lattices, it should be possible to improve the conjecture also for regular lattices.
Work in this direction is in progress.
\begin{acknowledgments}
One of the authors M. O. would like to thank hospitality of and by the members of Ko\c{c} University and Feza G\"ursey Institute in Turkey, acknowledges Dr. M. Hinczewski of Feza G\"ursey Institute for useful and fruitful discussions, Mr. S. Morita, and to Mr. Y. Matsuda of Tokyo Institute of Technology for many discussions and comments.
This work was partially supported by CREST, JST, by the 21st Century COE Program at Tokyo Institute of Technology `Nanometer-Scale Quantum Physics', and by the Grant-in-Aid for Scientific Research on the Priority Area ``Deepening and Expansion of Statistical Mechanical Informatics'' by the Ministry of Education, Culture, Sports, Science and Technology.
\end{acknowledgments}


\begin{thebibliography}{99}
\bibitem{EA}S. F. Edwards and P. W. Anderson, J. Phys. F {\bf 5}, 965 (1975).
\bibitem{SK}D. Sherrington and S. Kirkpatrick, Phys. Rev. Lett. {\bf 35}, 1792 (1975).
\bibitem{Young}A. P. Young (ed), {\em Spin Glasses and Random Fields} (World Scientific, Singapore, 1997).

\bibitem{HN81}H. Nishimori, Prog. Theor. Phys. {\bf 66}, 1169 (1981).
\bibitem{HNbook}H. Nishimori, {\em Statistical Physics of Spin Glasses and Information Processing: An Introduction} (Oxford Univ. Press, Oxford, 2001).
\bibitem{DoussalHarris}P. Le Doussal and A. B. Harris, Phys. Rev. Lett. {\bf 61}, 625 (1988).
\bibitem{NN}H. Nishimori and K. Nemoto, J. Phys. Soc. Jpn. {\bf 71}, 1198 (2002).
\bibitem{MNN}J.-M. Maillard, K. Nemoto and H. Nishimori, J. Phys. A {\bf 36}, 9799 (2003).
\bibitem{Nstat}H. Nishimori, J. Stat. Phys. {\bf 126}, 977 (2007).
\bibitem{TSN}K. Takeda, T. Sasamoto and H. Nishimori, J. Phys. A {\bf 38}, 3751 (2005).
\bibitem{NO}H. Nishimori and M. Ohzeki: J. Phys. Soc. Jpn. {\bf 75}, 034004 (2006).

\bibitem{HB}M. Hinczewski and A. N. Berker, Phys. Rev. B {\bf 72}, 144402 (2005).
\bibitem{Nobre}F. D. Nobre, Phys. Rev. E {\bf 64}, 046108 (2001).
\bibitem{Qeiroz}S. L. A. de Queiroz, Phys. Rev. B {\bf 73}, 064410 (2006).
\bibitem{pm2}N. Ito and Y. Ozeki, Physica {\bf A321}, 262 (2003).
\bibitem{pm3}F. Merz and J. T. Chalker, Phys. Rev. B {\bf 65}, 054425 (2002).
\bibitem{pm4}A. Honecker, M. Picco and P. Pujol, Phys. Rev. Lett. {\bf 87}, 047201 (2001).
\bibitem{pm5}F. D. A. Aar\~ao Reis, S. L. A. de Queiroz and R.R. dos Santos, Phys. Rev. B, {\bf 60} 6740 (1999).
\bibitem{Gaussian}M. Picco and A. Honecker and P. Pujol, J. Stat. Mech. {\bf 65}, P09006 (2006).
\bibitem{BO}A.N. Berker and S. Ostlund, J. Phys. C {\bf 12}, 4961 (1979).
\bibitem{GK}R.B. Griffiths and M. Kaufman, Phys. Rev. B {\bf 26}, 5022R (1982).
\bibitem{KG}M. Kaufman and R.B. Griffiths, Phys. Rev. B {\bf 30}, 244 (1984).
\bibitem{HB2}M. Hinczewski and A.N. Berker, Phys. Rev. E {\bf 73}, 066126 (2006).
\bibitem{MB}G. Migliorini and A.N. Berker, Phys. Rev. B {\bf 57}, 426 (1998).
\bibitem{Domany}E. Domany, J. Phys. C {\bf 12}, L119 (1979).
\end{thebibliography}
\end{document}